\documentclass[10pt,a4paper,twoside,twocolumn,american,prd,nofootinbib,superscriptaddress]{revtex4-1}
\usepackage{lmodern}

\usepackage[T1]{fontenc}
\usepackage[utf8]{inputenc}
\usepackage{graphicx}
\usepackage{color}
\usepackage{babel}
\usepackage{booktabs}
\usepackage{amsmath}
\usepackage{amssymb}
\usepackage{mathtools}
\usepackage[unicode=true,pdfusetitle,
 bookmarks=true,bookmarksnumbered=false,bookmarksopen=false,
 breaklinks=false,pdfborder={0 0 0},backref=false,colorlinks=true]
 {hyperref}
\hypersetup{
 citecolor=blue,filecolor=blue,linkcolor=blue,urlcolor=blue}

\makeatletter  

 \@ifundefined{textcolor}{}
 {%
   \definecolor{BLACK}{gray}{0}
   \definecolor{WHITE}{gray}{1}
   \definecolor{RED}{rgb}{1,0,0}
   \definecolor{GREEN}{rgb}{0,1,0}
   \definecolor{BLUE}{rgb}{0,0,1}
   \definecolor{CYAN}{cmyk}{1,0,0,0}
   \definecolor{MAGENTA}{cmyk}{0,1,0,0}
   \definecolor{YELLOW}{cmyk}{0,0,1,0}
 }

\makeatother 

\begin{document}

\title{Dark Matter Halo Sparsity of Modified Gravity Scenarios}

\author{Pier Stefano Corasaniti}
\email{Pier-Stefano.Corasaniti@obspm.fr}
\affiliation{LUTH, UMR 8102 CNRS, Observatoire de Paris, PSL Research University, Universit\'e Paris Diderot, 5 place Jules Janssen, 92195 Meudon, France}
\affiliation{Sorbonne Universit\'e, CNRS, UMR 7095, Institut d'Astrophysique de Paris, 98 bis bd Arago, 75014 Paris, France}
\author{Carlo Giocoli}
\affiliation{Dipartimento di Fisica e Astronomia, Alma Mater Studiorum Universit`a di Bologna, via Gobetti 93/2, I-40129 Bologna, Italy}
\affiliation{INAF, Osservatorio di Astrofisica e Scienza dello Spazio di Bologna, via Gobetti 93/3, I-40129 Bologna, Italy}
\affiliation{INFN, Sezione di Bologna, viale Berti Pichat 6/2, I-40127 Bologna, Italy}
\author{Marco Baldi}
\affiliation{Dipartimento di Fisica e Astronomia, Alma Mater Studiorum Universit`a di Bologna, via Gobetti 93/2, I-40129 Bologna, Italy}
\affiliation{INAF, Osservatorio di Astrofisica e Scienza dello Spazio di Bologna, via Gobetti 93/3, I-40129 Bologna, Italy}
\affiliation{INFN, Sezione di Bologna, viale Berti Pichat 6/2, I-40127 Bologna, Italy}

\begin{abstract}
Modified Gravity (MG) scenarios have been advocated to account for the dark energy phenomenon
in the universe. These models predict departures from General Relativity on large cosmic scales that can be tested through a variety of probes such as observations of galaxy clusters among others. Here, we investigate the imprint of MG models on the internal mass distribution of cluster-like halos as probed by the dark matter halo sparsity. To this purpose we perform a comparative analysis of the properties of the halo sparsity using N-body simulation halo catalogs of a standard flat $\Lambda$CDM model and MG scenarios from the DUSTGRAIN-{\it pathfinder} simulation suite. We find that the onset of the screening mechanism leaves a distinct signature in the redshift evolution of the ensemble average halos sparsity. Measurements of the sparsity of galaxy clusters from currently available mass estimates are unable to test MG models due to the large uncertainties on the cluster masses. We show that this should be possible in the future provided large cluster samples with cluster masses determined to better than $30\%$ accuracy level.  
\end{abstract}
\maketitle

\section{Introduction}\label{sec:intro}
Modified Gravity (MG) models have been proposed to explain the origin of the dark energy in the universe (for review, see e.g. \cite{Clifton2012,Austin2016}). In such scenarios departures from Einstein's theory of General Relativity (GR) occur at large cosmic scales resulting in a late-time phase of cosmic accelerated expansion. In contrast, GR is recovered at small scales through screening mechanisms that grant MG models to satisfy the stringent gravity constraints from Solar System experiments (see \cite{Khoury2010} for a general review). The deviations from GR at large scales not only modify the cosmic expansion, but also affect the growth of matter density fluctuations, thus leaving testable imprints on the cosmic structures (see \cite{Ishak2018,Ferreira2019} for a review of cosmological tests of MG). A particularly popular model of MG is the $f(R)$ gravity theory \cite{Buchdahl1970,Starobinski1980}, where the scalar curvature $R$ in the standard Einstein-Hilbert action integral is replaced by a function $f(R)$. More specifically, Hu \& Sawicki \cite{HuSawicki2007} have proposed a form of the $f(R)$ function which results in a cosmic background expansion that matches the standard $\Lambda $CDM one, though leaving direct imprints on the formation and evolution of cosmic structures.

Several studies have investigated the signature of MG models, and in particular of the Hu \& Sawicki \cite{HuSawicki2007} $f(R)$ gravity scenario, on galaxy cluster observables such as the cluster abundance and its redshift evolution \cite{Mak2012,Kopp2013,Achitouv2014,Cataneo2016}. Differences between the cluster mass profile inferred from lensing shear measurements and the dynamical analyses of the cluster content can also be indicative of violation of GR (see e.g. \cite{Terukina2014,Pizzuti2016}). This is because lensed photons tracing the mass profile do not experience fifth-force effects. This is not the case for dynamical mass estimates obtained from the analysis of galaxy dispersion velocities (see e.g. \cite{Pizzuti2017}) or the hydrostatic equilibrium of the intra-cluster gas \cite{Winther2012}. Moreover, fifth-force effects can modify the cluster mass-temperature relation \cite{Hammami2017}. Finally, the growth of structures in MG scenarios can lead to halo mass profiles which differ from those expected in the standard $\Lambda$CDM model (see e.g. \cite{Schmidt2009,Hammami2015,Achitouv2016}).

Here, we study the imprint of $f(R)$ MG models on the sparsity of dark matter halos. Originally introduced in \cite{Balmes2014}, the halo sparsity provides a direct observational proxy of the mass distribution in halos. Its use as cosmic probe has been extensively investigated in the literature \cite{Balmes2014,Corasaniti2018,Corasaniti2019}. These studies have shown that the halo sparsity provides several advantages compared to testing cosmology with the more common concentration-mass relation approach (see e.g. \cite{Ettori2010}). On the one hand, it is much less affected by astrophysical systematics as well as selection effects. On the other hand its ensemble average value at a given redshift can be predicted from prior knowledge of the halo mass function. This provides a simple quantitative framework to perform cosmological parameter inference analyses. Past studies have mainly investigated the basic properties of the halo sparsity in the context of $\Lambda$CDM-like models. The work presented here intends to extend these analyses to MG scenarios. To this purpose we have performed a numerical study using N-body halo catalogs of $f(R)$ model simulations. We find that the basic properties of the halo sparsity hold valid also in the case of MG models. Furthermore, the different onset of the screening mechanism leaves a distinct imprint of the $f(R)$ models on the redshift evolution of the ensemble average halo sparsity. We show that this can be tested with measurements of the sparsity of galaxy clusters. We perform a simple data model comparison as working example and evaluate the observational requirements that in the future would allow to distinguish among the simulated models.

The paper is organised as follows. In Section~\ref{methodology} we describe the numerical simulation dataset and review the basic properties of the halo sparsity. We present the results of our study in Section~\ref{results} and Section~\ref{forecast}, while in Section~\ref{conclu} we present our conclusions.

\section{Methodology}\label{methodology}

\subsection{Dark Matter Halo Sparsity Primer}
The dark matter halo sparsity is defined as \citep{Balmes2014}:
\begin{equation}
s_{\Delta_1,\Delta_2} = \frac{M_{\Delta_1}}{M_{\Delta_2}},
\end{equation}
where $M_{\Delta_1}$ and $M_{\Delta_2}$ are the halo masses enclosing the overdensities $\Delta_1$ and $\Delta_2$ respectively with $\Delta_1<\Delta_2$ (in units of the background density $\rho_b$ or the critical density $\rho_c$). This provides a non-parametric characterisation of the halo mass distribution, while carrying cosmological information encoded in the mass profile. As shown in \cite{Balmes2014}, the properties of the halo sparsities are independent of the choice of the overdensity units. Moreover, the cosmological signal encoded in the sparsity increases as the difference between $\Delta_1$ and $\Delta_2$ is largest. Nonetheless, the values of $\Delta_1$ and $\Delta_2$ cannot be arbitrarily chosen. On the one hand values of $\Delta_1\lesssim 100$ should be excluded since the definition of halo as a distinct object becomes ambiguous, while for $\Delta_2\gtrsim 2000$ the sparsity probes inner regions of the halo mass profile where astrophysical processes acting on the baryon content may alter the halo mass distribution and dilute the cosmological signal. 

Halos with density radial distributions which are well fit by the Navarro-Frenk-White profile \cite{NFW1997} have sparsity values that are in a one-to-one correspondence with the value of the concentration parameter. In such a case the sparsity does not provide additional information on the halo mass distribution compared to that already encoded in the concentration. However, as shown in \cite{Balmes2014} not all halos have profiles that are exactly described by the NFW function. Consequently, the concentration is no longer informative of the halo density profile and its cosmological dependence. This is not the case for the halo sparsity which remains close to a constant value with a small intrinsic scatter even for halos which exhibit large deviations from the NFW profile. Such constant value is found to only vary with redshift and cosmology, thus providing a cosmological proxy. This can be easily understood by noticing that the sparsity quantifies the mass excess between the radii $r_{\Delta_1}$ and $r_{\Delta_2}$ relative to the mass enclosed in the inner radius $r_{\Delta_2}$, i.e. $s_{\Delta_1,\Delta_2}=\Delta{M}/M_{\Delta_2}+1$. Hence, at a given redshift the sparsity is smaller in a cosmological model where the assembly of a halo occurs at earlier times, since the mass assembled within the inner radius is greater than in a model where the halo formation is delayed \cite{Giocoli2007,Giocoli2012}.

As shown in \cite{Balmes2014,Corasaniti2018,Corasaniti2019} a key property of the halo sparsity is its nearly independence on the halo mass $M_{\Delta_1}$. In fact, this implies that the ensemble average sparsity at a given redshift can be predicted from prior knowledge of the halo mass function at the overdensity of interests. More specifically, it follows that:
\begin{equation}
\int_{{M}^{\rm min}_{\Delta_2}}^{{M}^{\rm max}_{\Delta_2}}\frac{dn}{d{M}_{\Delta_2}}d\ln{{M}_{\Delta_2}}= y  \int_{y \cdot {M}^{\rm min}_{\Delta_2}}^{y \cdot {M}^{\rm max}_{\Delta_2}}  \frac{dn}{d{M}_{\Delta_1}}d\ln{{M}_{\Delta_1}},\label{sparpred}
\end{equation}
which can be solved numerically to infer the value of $y=\langle s_{\Delta_1,\Delta_2}\rangle$ given the mass functions $dn/dM_{\Delta_1}$ and $dn/dM_{\Delta_2}$ respectively. The validity of Eq.~(\ref{sparpred}) has been extensively tested in \cite{Balmes2014,Corasaniti2018,Corasaniti2019}. It is worth noticing that the validity of Eq.~(\ref{sparpred}) also implies the validity of the following relation:
\begin{equation}\label{spars_rel}
\langle s_{\Delta_1,\Delta_2}\rangle \equiv \left\langle\frac{M_{\Delta_1}}{M_{\Delta_2}}\right\rangle \approx \frac{\langle 1/M_{\Delta_2}\rangle}{\langle 1/M_{\Delta_1}\rangle}.
\end{equation}
Hence, for an ensemble of halos one has three distinct ways to estimate the average halo sparsity which provide a set of consistency relations whose validity can be used to test the presence of outliers in galaxy cluster samples \cite{Corasaniti2019}.

Hereafter, we will test the validity of these properties in the context of MG models. 

\subsection{Numerical Simulation Dataset}
We use the numerical halo catalogs from the DUSTGRAIN-{\it pathfinder} simulation suite \cite{Giocoli2018}. This consists of N-body simulations of a ($750$ Mpc/h)$^3$ volume with $N_p=768^3$ particles (corresponding to a mass resolution of $m_p\approx 8\cdot 10^{10}\,M_{\odot}/h$) of a flat $\Lambda$CDM scenario and three MG models (with and without massive neutrinos) in the form of the Hu \& Sawicki \cite{HuSawicki2007} $f(R)$ gravity theory. In particular, the $f(R)$ gravity models included in our simulation suite are specified by the following values of the characteristic parameter $f_{R0}=-10^{-4}$ ($fR4$), $-10^{-5}$ ($fR5$) and $-10^{-6}$ ($fR6$). The standard cosmological parameters have been set to values consistent with the results of the {\it Planck}-2015 cosmological data analysis \cite{Planck2015}: matter density $\Omega_m=0.31345$, baryon density $\Omega_b=0.0481$, Hubble constant $H_0=67.31$ km s$^{-1}$ Mpc$^{-1}$, scalar spectral index $n_s=0.9658$ and root-mean-square amplitude of the linear density fluctuations on the $8$ Mpc/h scale $\sigma_8=0.847$. The simulations have been carried out with the \small{MG-Gadget} code \cite{PuchweinBaldiSpringel2013}. We refer the readers to \cite{Giocoli2018} for a detailed description of the simulation characteristics. 

Here, we only consider the halo catalogs from the simulations without massive neutrinos. Halos in the simulations have been detected using the spherical overdensity algorithm \cite{Tormen2004,Giocoli2008,Despali2016} at overdensities $\Delta=200\rho_c,500\rho_c$ and $1000\rho_c$ respectively. Since we are interested on cluster-size halos we limit our analysis to halos with mass $M_{200c}>10^{13}M_{\odot}\,h^{-1}$. Moreover, we focus on a subsample of {\it matched halos}, i.e. identical halos in the different overdensity catalogs, such that the values of $M_{200c}$, $M_{500c}$ and $M_{1000c}$ concern the mass profile of the same halo. To this purpose, for each redshift snapshot we have identified all the halos in the $\Delta=500\rho_c$ and $1000\rho_c$ catalogs which have center-of-mass coordinates that differ from those of the halos in the $\Delta=200\rho_c$ catalog by less than the spatial resolution of the simulation.

\begin{figure*}[th]
\centerline{\includegraphics[width=1.85\columnwidth]{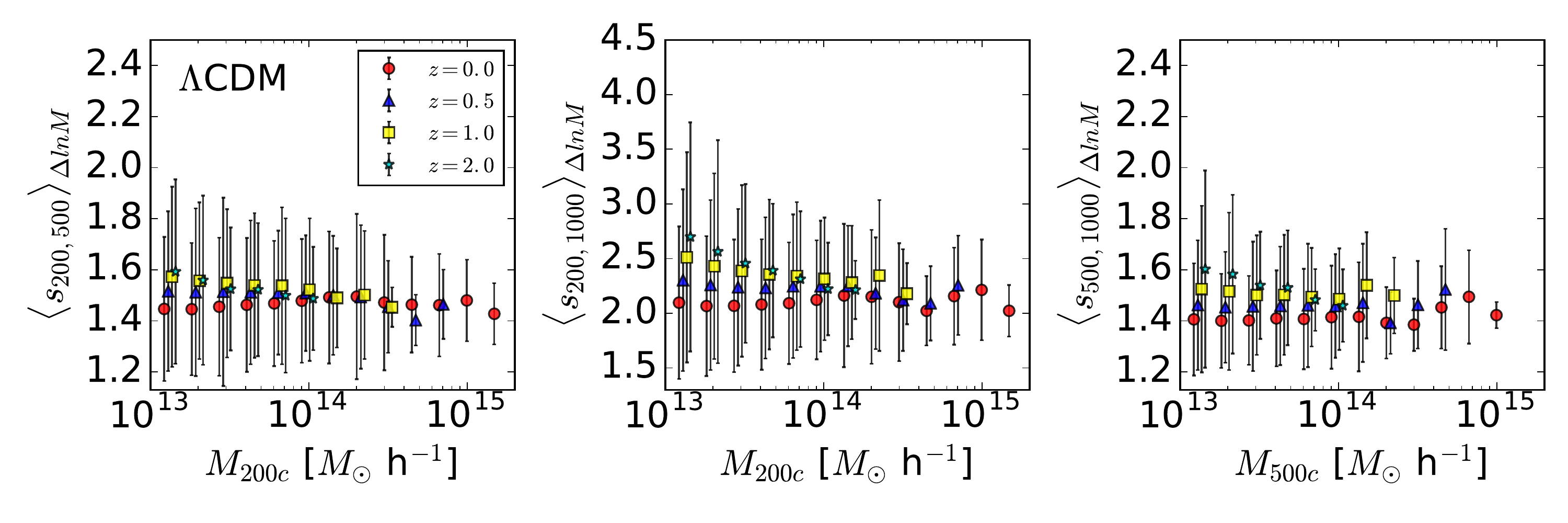}}
\caption{\label{fig:spars_LCDM} Average halo sparsity $\langle s_{200,500}\rangle$ (left panel), $\langle s_{200,1000}\rangle$ (middle panel) and $\langle s_{500,1000}\rangle$ as function of $M_{\Delta_1}$ in mass bins of size $\Delta\ln{M_{\Delta_1}}=0.3$ at $z=0$ (red circles), $0.5$ (blue triangles), $1$ (yellow squares) and $2$ (cyan stars) for the $\Lambda$CDM model simulation. The mass points at different redshifts have been displaced for visual purposes. The errors bars correspond to the standard deviation around the mean value, which is dominated by the intrinsic scatter of the halo sparsity.}
\end{figure*}

\begin{figure*}[th]
\centerline{\includegraphics[width=1.85\columnwidth]{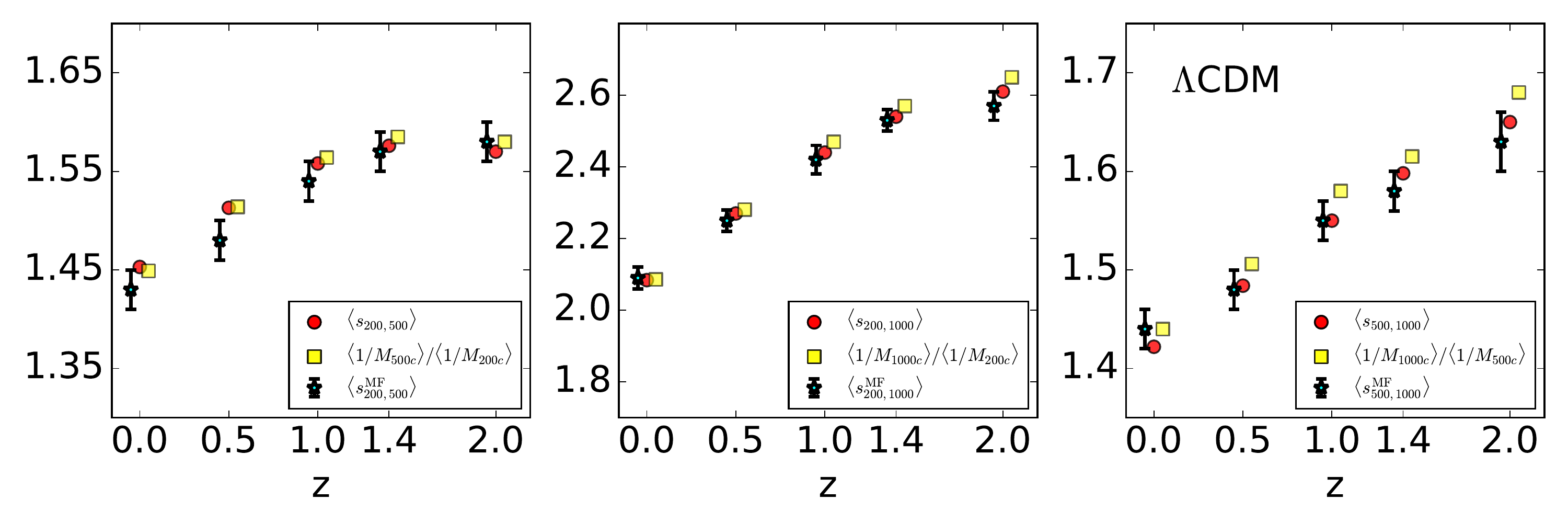}}
\caption{\label{fig:spars_rel_LCDM} Ensemble average sparsity estimates for the $\Lambda$CDM model. Red circles corresponds to the ensemble average halo sparsity $\langle s_{200,500}\rangle$ (left panel), $\langle s_{200,1000}\rangle$ (middle panel) and $\langle s_{500,1000}\rangle$ at $z=0,0.5,1,1.4$ and $2$ respectively. In each panel the yellow squares corresponds to the estimate of the average sparsity from Eq.~(\ref{spars_rel}), while the cyan stars are the estimates from the mass function relation Eq.~(\ref{sparpred}).}
\end{figure*}

\section{Results}\label{results}
\subsection{Standard $\Lambda$CDM Model}
We compute the halo sparsity $s_{200,500}$, $s_{200,1000}$ and $s_{500,1000}$ of the matched halos from the $\Lambda$CDM model simulation. 
In Fig.~\ref{fig:spars_LCDM} we plot the averaged sparsities as function of $M_{\Delta_1}$ in logarithmic mass bins of size $\Delta\ln{M_{\Delta_1}}=0.3$ at different redshifts. The error bars represent the dispersion around the mean which is dominated by the intrinsic scatter of the halo sparsity. As expected, we find that the average sparsity remains constant as function of halo mass to very good approximation and well within the estimated dispersion. In all cases, the variation slightly increases with redshift, though never exceeding the $6\%$ level over two decades in mass at $z=2$. The dispersion also remains roughly constant with halo mass and of order of $\sim 20\%$ level. This is consistent with the findings of \cite{Balmes2014,Corasaniti2018}. 
In Fig.~\ref{fig:spars_rel_LCDM} we plot the values of the ensemble average halo sparsities at different redshifts against the values inferred from the halo mass function relation Eq.~(\ref{sparpred}) and the evaluation of the ratio of the ensemble averages of the inverse halo masses as given by Eq.~(\ref{spars_rel}). In order to compute Eq.~(\ref{sparpred}) we have used a polynomial fit to the numerical halo mass function of the matched halos at the different overdensities. The uncertainties on the estimated value of the average sparsity correspond to the propagation of the Poisson errors on the estimated mass functions. We can see that also these estimates agree to numerical accuracy with one another, which is consistent with the findings of \cite{Corasaniti2019}. We would like to stress that the redshift dependence of the sparsity in the interval $0<z<1$ shown in Fig.~\ref{fig:spars_rel_LCDM} is not a spurious effect of the pseudo-evolution of the halo mass due to the variation of the reference density \cite{Diemer2013}. In fact, being the sparsity a mass ratio the pseudo-evolution between the two overdensities cancels out at leading order, while it is the difference in the physical growth of the halo that drives the redshift variation of the sparsity.

\begin{figure*}[th]
\centerline{\includegraphics[width=1.85\columnwidth]{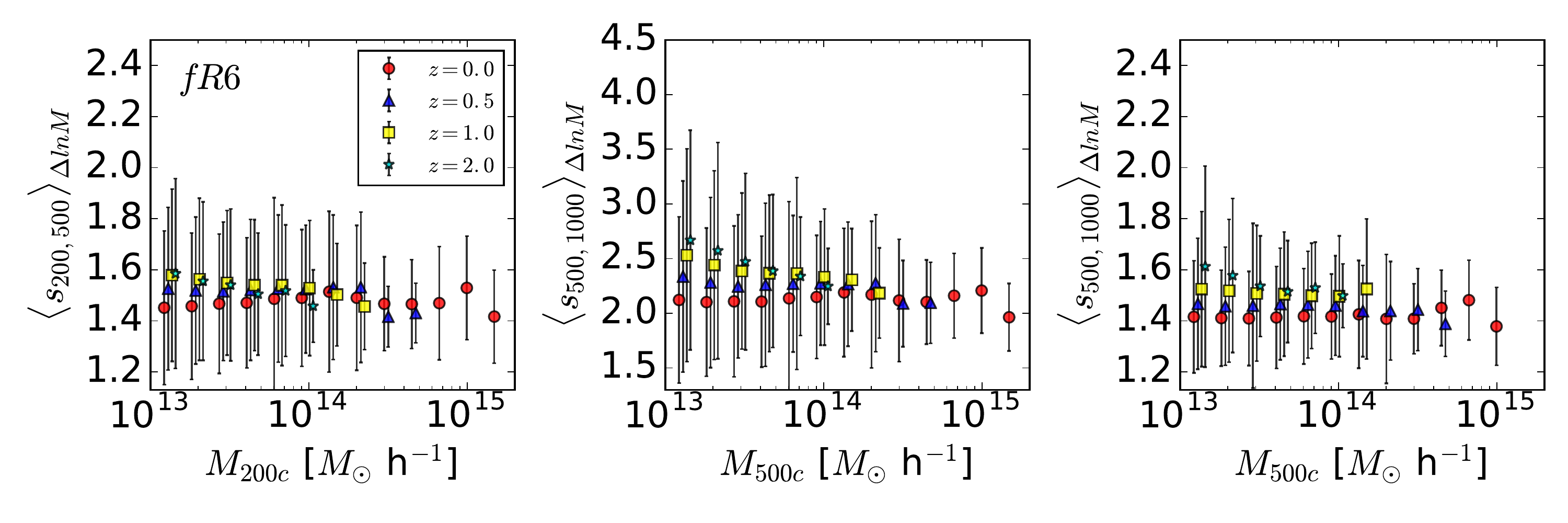}}
\centerline{\includegraphics[width=1.85\columnwidth]{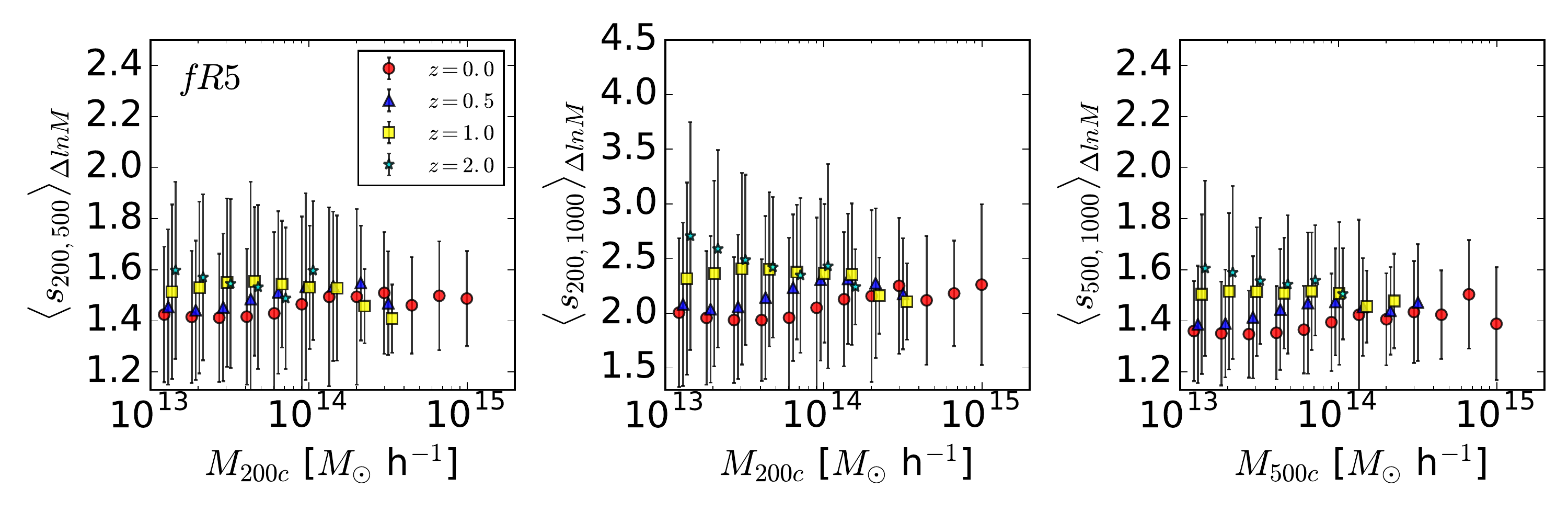}}
\centerline{\includegraphics[width=1.85\columnwidth]{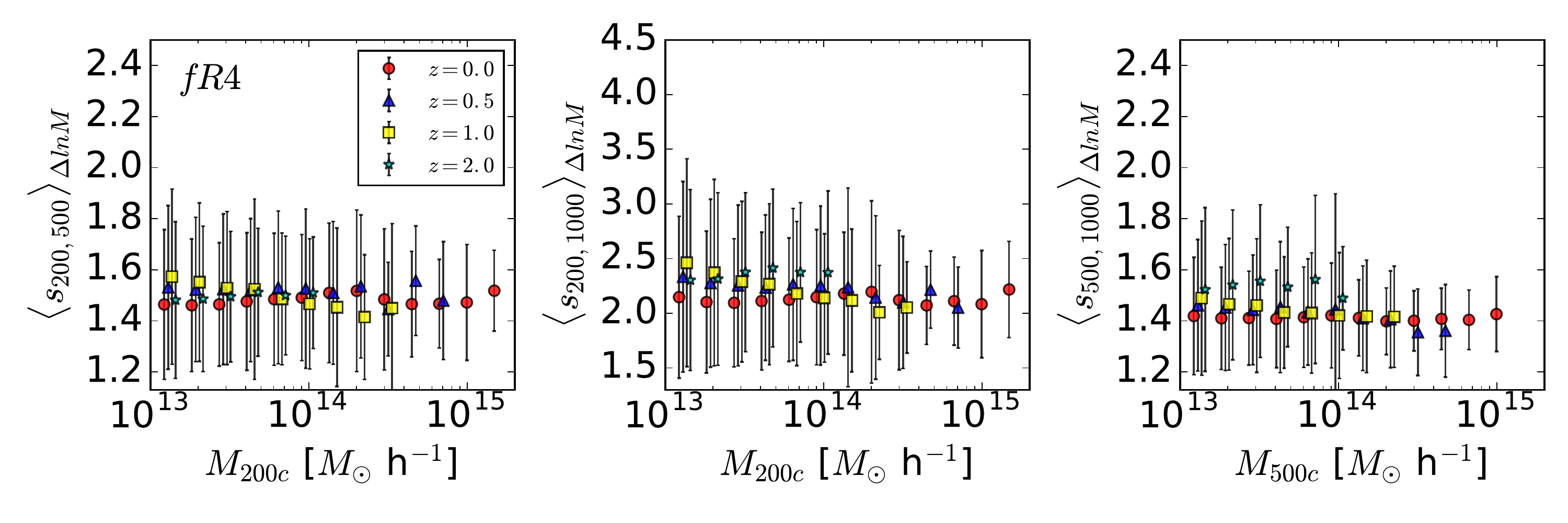}}
\caption{\label{fig:spars_fR} As in Fig.~\ref{fig:spars_LCDM} for the MG models: $fR6$ (top panels), $fR5$ (central panels) and $fR4$ (bottom panels) respectively.}
\end{figure*}

\begin{figure*}[tb]
\centerline{\includegraphics[width=1.85\columnwidth]{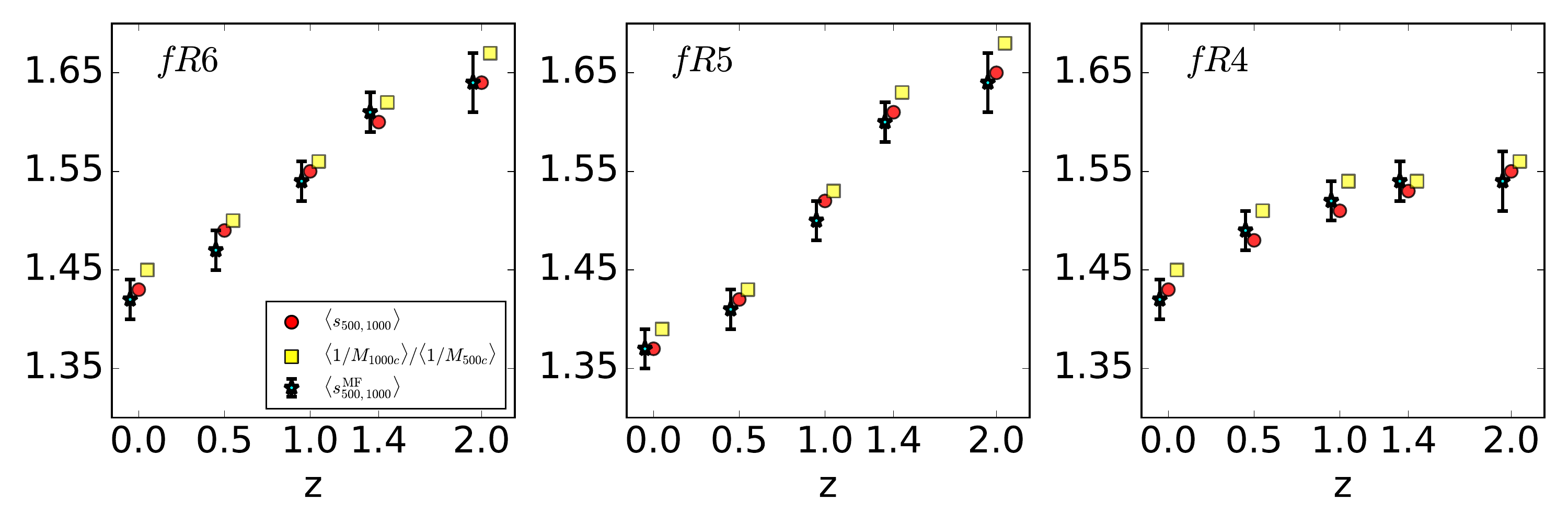}}
\caption{\label{fig:spars_rel_MG} Ensemble average halo sparsity $\langle s_{500,1000}\rangle$ estimates for $fR6$ (left panel), $fR5$ (central panel) and $fR4$ (right panel) at $z=0,0.5,1,1.4$ and $2$ respectively.}
\end{figure*}

\begin{figure*}[tb]
\centerline{\includegraphics[width=1.9\columnwidth]{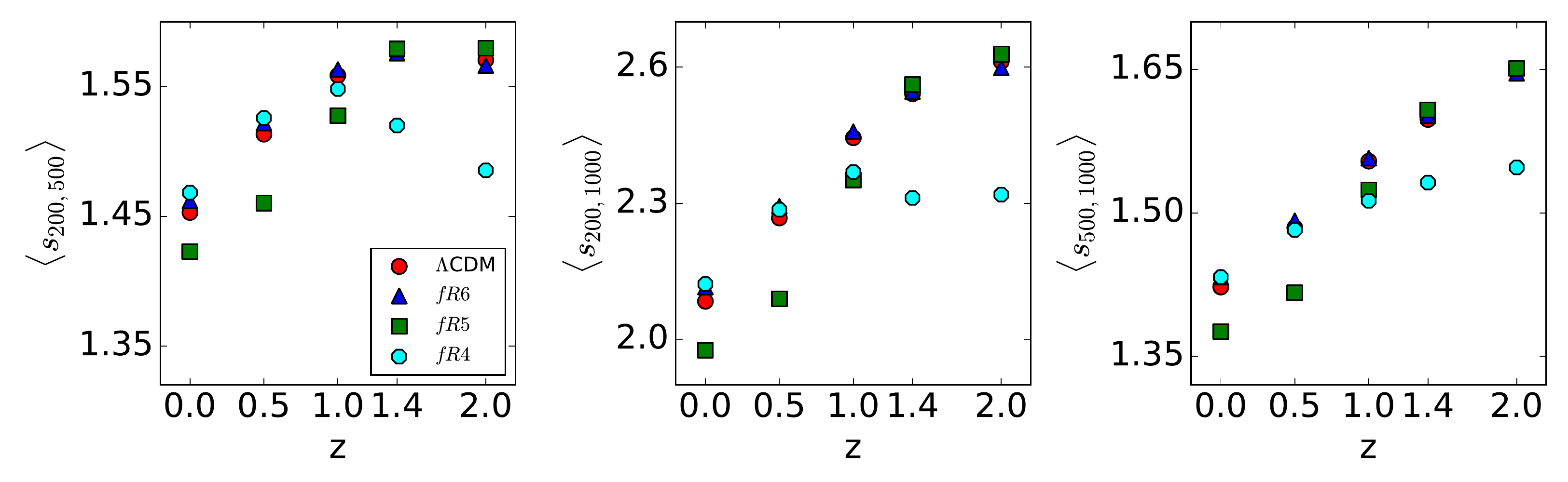}}
\caption{\label{fig:spars_cosmo} Redshift evolution of the ensemble average halo sparsities $\langle s_{200,500}\rangle$ (left panel), $\langle s_{200,1000}\rangle$ (middle panel) and $\langle s_{500,1000}\rangle$ for $\Lambda$CDM (red circles), $fR6$ (blue triangles), $fR5$ (green squares) and $fR4$ (cyan circles) respectively.}
\end{figure*}

\subsection{$f(R)$ Models}
Similarly to the $\Lambda$CDM case, we compute the halo sparsity $s_{200,500}$, $s_{200,1000}$ and $s_{500,1000}$ of the matched halos from the $f(R)$ model simulations. 
In Fig.~\ref{fig:spars_fR} we plot the average sparsities as function of $M_{\Delta_1}$ in mass bins of size $\Delta\ln{M_{\Delta_1}}=0.3$ at $z=0$ (red circles), $0.5$ (blue triangles), $1$ (yellow squares) and $2$ (cyan stars) for the $fR6$ (top panels), $fR5$ (central panels) and $fR4$ (bottom panels) models respectively. As in the $\Lambda$CDM case, we can see that the halo sparsity remains constant to good approximation as function of $M_{\Delta_1}$. Notice that the $fR6$ case exhibits a trend which closely matches that of the $\Lambda$CDM model. This is not surprising since in the $fR6$ model the deviations from GR occurs on such large scales that even the most massive halos are screened. In Fig.~\ref{fig:spars_rel_MG} we plot the comparison between the ensemble average halo sparsity $\langle s_{500,1000}\rangle$ against the estimates obtained from Eq.~(\ref{sparpred}) and Eq.~(\ref{spars_rel}) for the different MG models. Again, we find the different average sparsity estimates to be consistent with one another. For conciseness we do not show the other sparsity configurations for which we find such consistency relations to also stand true. Notice that the validity of Eq.~(\ref{sparpred}) implies that the availability of a parametric modelling of the halo mass function for different mass overdensities can provide us with a viable tool to predicted the average halo sparsity for a generic MG model. This is an aspect that is key to perform parameter inference analysis of MG models from sparsity measurements and which we leave to future work.

The fact that the basic sparsity properties hold valid also in the case of MG models is essentially because the assembly of dark matter halos remains a bottom-up process. Differences with the respect to the $\Lambda$CDM scenario only manifest at the level of the redshift evolution of the ensemble average halo sparsity. This can be better appreciated in Fig.~\ref{fig:spars_cosmo} where we plot $\langle s_{200,500}\rangle$, $\langle s_{200,1000}\rangle$ and $\langle s_{500,1000}\rangle$ as function of redshift. Notice that the evolution of average sparsities in the $fR6$ model closely matches that of the $\Lambda$CDM, this is not the case of the $fR5$ and $fR4$ models which depart from the $\Lambda$CDM prediction at low and high redshift respectively. These trends result from the different onset of the screening mechanism in the simulated MG models and the effect of the fifth-force on the halo formation. As clearly shown in \cite{Li2013}, the larger the deviations from GR (i.e. the greater $|f_{R0}|$) the earlier the onset of the fifth-force, this has the effect of increasing the growth of matter density fluctuations from small to large scales relative to the $\Lambda$CDM model. Because of this, dark matter halos assemble at earlier time than in $\Lambda$CDM, consequently the inner halo mass is larger (i.e. the halo is more concentrated) and the corresponding sparsity is lower. Once the massive halos are assembled, the impact on the halo mass profile depends on the mass scale of the screening mechanism. As an example, the authors of \cite{Hagstotz2019} have investigate the imprint of the screening mechanism on the velocity dispersion of massive halos using the same DUSTGRAIN-{\it pathfinder} suite catalogs. They have found that in the $fR6$ model all relevant halo mass scales are screened, consequently the velocity dispersion in cluster-size halos matches that of the $\Lambda$CDM model. In contrast, for the $fR4$ model all mass scales are unscreened and the velocity dispersions are boosted by a constant factor as function of halo mass. The case $fR5$ represents an intermediate situation with the fifth-force enhancing by a constant factor the velocity dispersion for masses $\lesssim 10^{13}-10^{14} M_{\odot}$ h$^{-1}$. Such effects can account for the trends shown in Fig.~\ref{fig:spars_cosmo}. More specifically, in the $fR4$ model the growth of matter density fluctuations is larger at earlier times than in the $\Lambda$CDM case. This leads to the formation of more concentrated halos at high-redshifts, thus resulting in a lower average sparsity than in $\Lambda$CDM. However, once the most massive halos are assembled at later times, the fifth-force effect equally enhance their mass profiles at different radii consistently with the finding of \cite{Hagstotz2019} which results in an average sparsity evolution that follows that of the standard $\Lambda$CDM scenario. In the $fR5$ case the emergence of fifth-force effects occur at later times than $fR4$. Hence, while at high-redshift the halo formation is similar to the $\Lambda$CDM case, at low-redshifts the halos are more concentrated, thus resulting in lower average halo sparsities.

The trends shown in Fig~\ref{fig:spars_cosmo} suggest that an accurate determination of the average sparsity in large samples of clusters at different redshifts can provide a complementary test of the imprints MG models.

\section{Testing Modified Gravity with Galaxy Cluster Sparsity}\label{forecast}
The sparsity of galaxy clusters can be determined from cluster mass measurements at different overdensities. These can be obtained through a variety of methods (see e.g. \cite{Pratt2019} for a review). Mass measurements based on the analysis of shear lensing profile of clusters give estimates that are to a large extent independent of the specificities of the MG models considered. This is because in a wide range of MG scenarios lensed photons are not affected by fifth-force effects. Hence, the inferred cluster sparsities can be directly compared to that predicted from the analysis of N-body halo catalogs.

As shown in \cite{Corasaniti2018}, measurements of the cluster sparsity can be used to infer cosmological parameter constraints. The halo sparsity is primarily sensitive to the cosmic matter density $\Omega_m$ and the amplitude of matter density fluctuations $\sigma_8$. These are degenerate parameters since they both determine the overall amplitude and redshift evolution of the average halo sparsity. Because of this, cluster sparsities mainly constrain the combination $S_8=\sigma_8\sqrt{\Omega_m}$. As we have seen in the previous section, MG models introduce a characteristic imprint on the redshift evolution of the average sparsity. Hence, we may expect that the constraints on the MG amplitude parameter $|f_{R0}|$ are less affected by degeneracies with other cosmological parameters such as $S_8$, though they may impact the overall goodness-of-fit. 

\begin{figure}[t]
\centerline{\includegraphics[width=0.8\columnwidth]{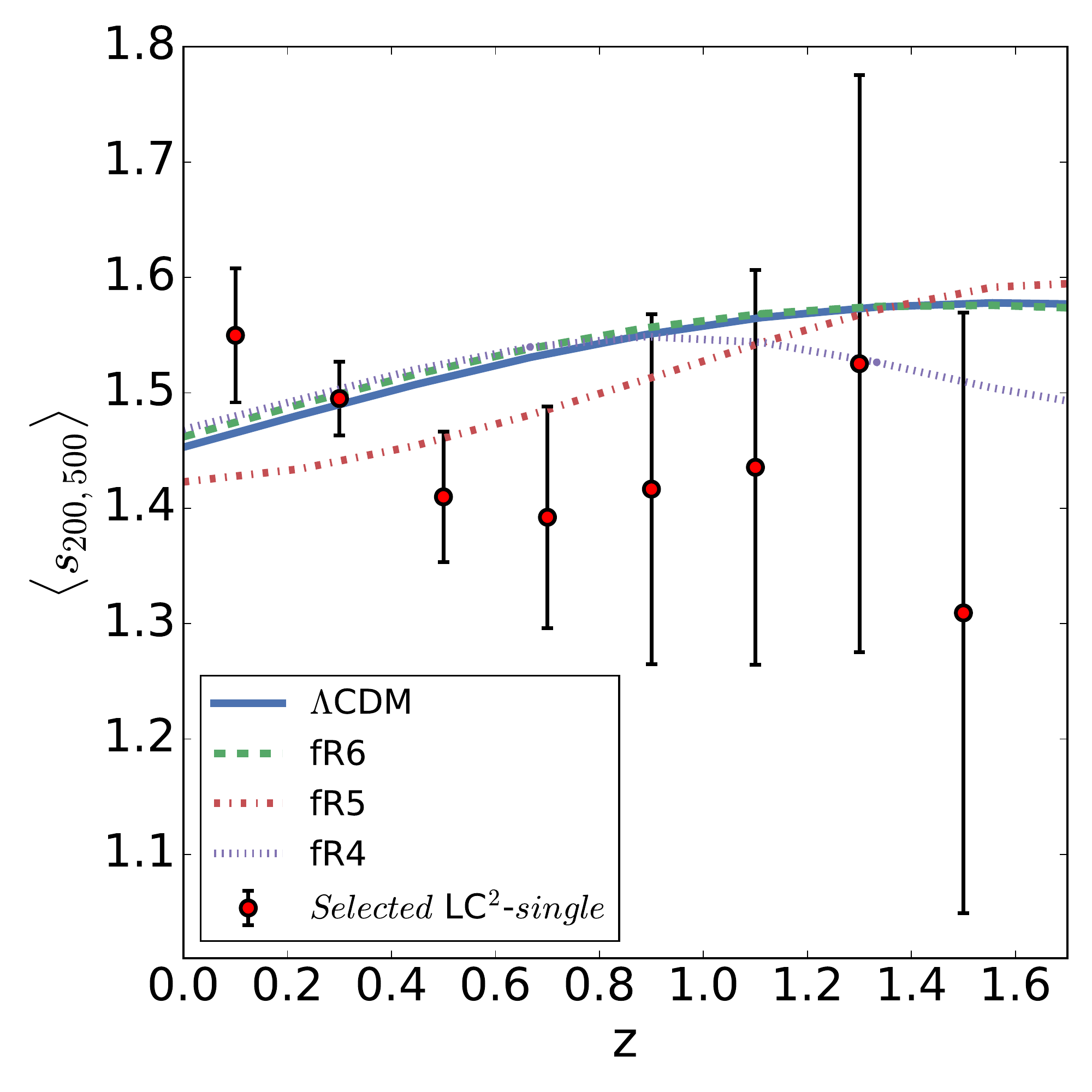}}
\caption{\label{fig:spars_lc2single} Interpolated redshift evolution of the ensemble average halo sparsities $\langle s_{200,500}\rangle$ for $\Lambda$CDM (blue solid line), $fR6$ (green dashed line), $fR5$ (red dash-dot line) and $fR4$ (magenta dot line) from the DUSTGRAIN-{\it pathfinder} matched halo catalogs against average cluster sparsity estimates in redshift bins of size $\Delta{z}=0.2$ from a selected sample of lensing mass measurements from the LC$^2$-{\it single} catalog \cite{Sereno2015}.}
\end{figure} 

\begin{figure}[th]
\centerline{\includegraphics[width=0.8\columnwidth]{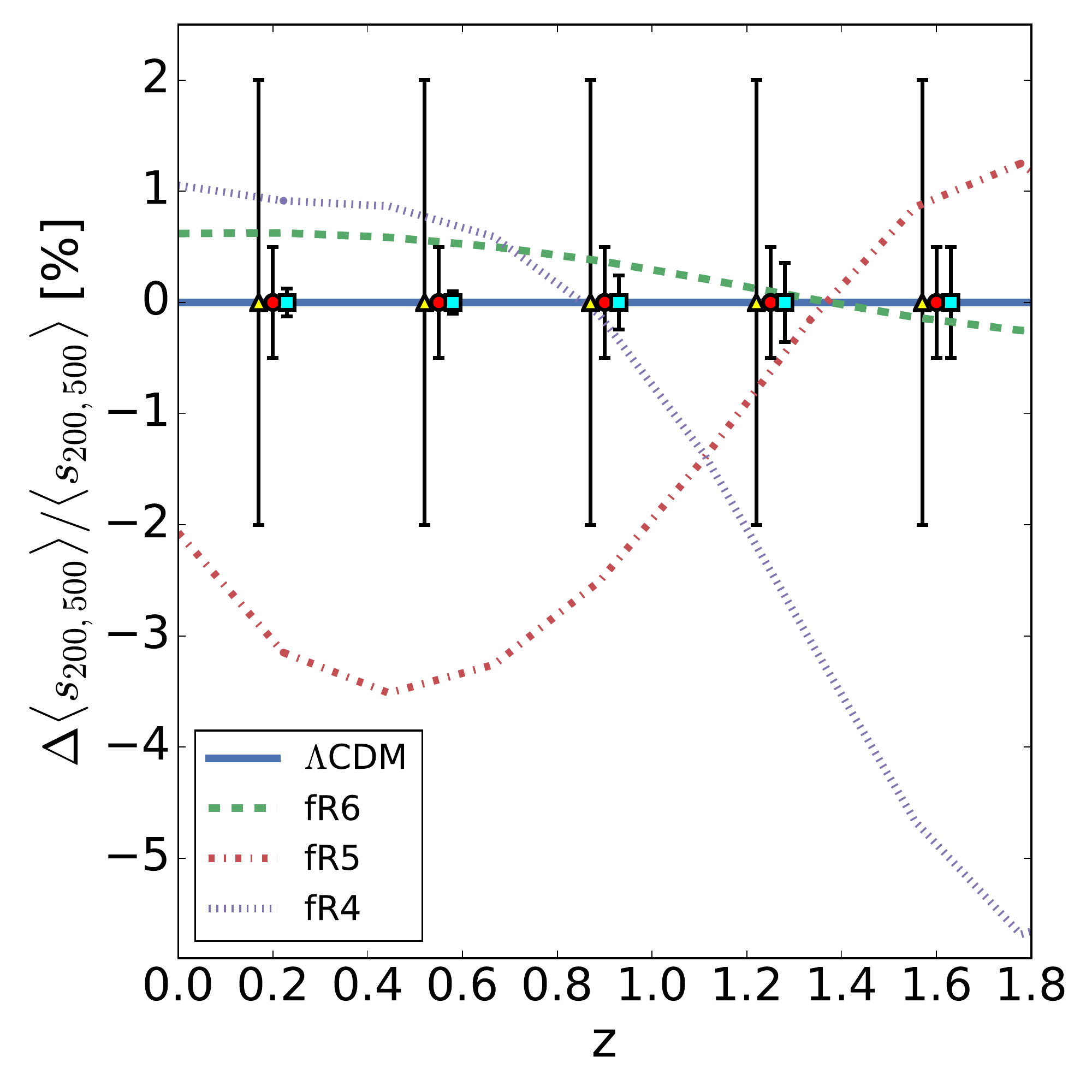}}
\caption{\label{fig:spars_forecast} Relative difference of the average sparsity redshift evolution with respect to the $\Lambda$CDM case (blue solid line) for $fR6$ (green dashed line), $fR5$ (red dash-dot line) and $fR4$ (magenta dot line) against forecasted data assuming $1000$ clusters equally distributed in $5$ redshift bins in the range $0<z<1.6$ with $30\%$ (yellow triangles) and $5\%$ (red circles) fractional mass errors respectively; a sample from a Euclid-like survey of several hundred clusters with mass errors $\lesssim 5\%$ up to $z=1.6$ as in \cite{Sartoris2016} (cyan squares).}
\end{figure} 

It is beyond the scope of this work to perform a full cosmological parameter inference analysis of MG models. Nevertheless, as a working example we compare the redshift evolution of the average sparsity $\langle s_{200,500}\rangle$ from the DUSTGRAIN-{\it pathfinder} simulations against estimates from a selected sample of lensing shear profile masses from the Literature Catalogs Lensing Clusters\footnote{\url{http://pico.oabo. inaf.it/~sereno/CoMaLit/LC2/}} (LC$^2$-{\it single}, \cite{Sereno2015}). More specifically, we focus on cluster mass measurements obtained from a 2-parameter fit of the measured shear profile of clusters. For each of the 187 clusters in our sample we compute the sparsity from the ratio of the available mass measurements at $\Delta=200\rho_c$ and $500\rho_c$ and estimate the uncertainties from error propagation of the mass measurement errors. We evaluate the average sparsity in different redshift bins of size $\Delta{z}=0.2$. These are shown in Fig.~\ref{fig:spars_lc2single} against the interpolated trends from the N-body halo samples. It is worth remarking that the uncertainties on the average sparsity estimates are dominated by the errors on the cluster mass measurements. We have compared the curves of the different models against the data to find the following $\chi^{2}$ values: $\chi^2_{\Lambda{\rm CDM}}=10.1$, $\chi^2_{fR6}=10.6$, $\chi^2_{fR5}=11.1$ and $\chi^2_{fR4}=9.9$. As we can see the differences in the goodness-of-fit are $\Delta{\chi^2}\lesssim 1.2$, thus the models are statistically indistinguishable from one another with current cluster sparsity measurements.

It is instructive to estimate the level of accuracy necessary for average sparsity measurements to distinguish the MG models considered here. Following \cite{Corasaniti2018}, we model the errors on the average sparsity at a given redshift as:
\begin{equation}
\sigma_{z}=\langle s^{\rm fid}_{200,500}(z)\rangle e_{M}\sqrt{2/N(z)},
\end{equation}
where $\langle s^{\rm fid}_{200,500}(z)\rangle$ is the average sparsity of the fiducial cosmological model, $e_{M}$ is the fractional mass and $N(z)$ is the number of clusters at redshift $z$. 

In Fig.~\ref{fig:spars_forecast} we plot the difference of the average sparsity trends shown in Fig.~\ref{fig:spars_lc2single} relative to the $\Lambda$CDM model against expectations for three different observational scenarios consisting of a sample of $1000$ clusters equally distributed in redshift bins in the range $0.2<z<1.6$ with fractional mass errors of $30\%$ (yellow triangles) and $5\%$ (red circles) respectively, and a sample from a Euclid-like survey (cyan square). In the latter case, we assume the redshift distribution of several hundred clusters with fractional errors on weak lensing estimated mass $\lesssim 5\%$ up to $z\sim 1.6$ as investigated in \cite{Sartoris2016}. We can see that cluster mass measurements with a $30\%$ accuracy would be able to distinguish the $fR4$ and $fR5$ models from $\Lambda$CDM, while a Euclid-like survey may be able to test deviations from GR at the level of $fR6$ models.

\section{Conclusions}\label{conclu}
The dark matter halo sparsity provides a non-parametric characterisation of the mass distribution of halos in terms of the ratio of the halo masses enclosing two different overdensities. This carries cosmological dependent information encoded in the halo mass profile and can be tested using mass estimates of galaxy clusters. Previous studies in the literature have investigated the properties of the halo sparsity and its use as a complementary probe of galaxy cluster cosmology in the context of $\Lambda$CDM-like models. Here, we have extended these analyses to the case of $f(R)$ MG scenarios. To this purpose we have used numerical N-body halo catalogs from the DUSTGRAIN-{\it pathfinder} simulations suite. We have shown that similarly to $\Lambda$CDM cosmologies, the halo sparsity remains nearly constant as function of halo mass with a relatively small intrinsic scatter. This implies that its ensemble average value can be inferred from prior knowledge of the halo mass function at the overdensity of interests. Another consequence of this property is the fact that the ensemble average sparsity coincides with the ratio of the ensemble average of the inverse halo masses at the overdensities considered. We have found both properties to be valid also for $f(R)$ MG models. Finally, we have found that $f(R)$ MG models leave a distinct imprint on the redshift evolution of the average halo sparsity that differs from that expected in $\Lambda$CDM. In particular, we have shown that the different redshift trends are a manifestation of the different onset of the screening mechanism in the simulated MG models. Average sparsity estimates from shear lensing mass measurements of galaxy clusters currently available are unable to distinguish among the predicted trends. However, this could be achieved in the future with sufficiently large cluster samples provided cluster masses are determined to better than $30\%$ accuracy level.

\begin{acknowledgments}
We are grateful to Mauro Sereno for providing the selected sample of cluster masses from the LC$^2$ \cite{Sereno2015}. PSC would like to thank Yann Rasera for useful comments. The DUSTGRAIN-{\it pathfinder} simulations discussed in this work have been performed and analysed on the Marconi supercomputing machine at Cineca thanks to the PRACE project SIMCODE1 (grant nr. 2016153604) and on the computing facilities of the Computational Center for Particle and Astrophysics (C2PAP) and of the Leibniz Supercomputer Center (LRZ) under the project ID pr94ji. CG and MB acknowledge support from the grants ASI n.I/023/12/0, ASI-INAF n. 2018-23-HH.0 and PRIN MIUR 2015 "Cosmology and Fundamental Physics: illuminating the Dark Universe with Euclid". CG acknowledges support from PRIN-MIUR 2017 WSCC32 ``Zooming into dark matter and proto-galaxies with massive lensing clusters''. MB acknowledges support from PRIN MIUR 2017 ``Combining Cosmic Microwave Background and Large Scale Structure data: an Integrated Approach for Addressing Fundamental Questions in Cosmology'' (2017YJYZAH). The research leading to these results has received funding from the European Research Council under the European Community Seventh Framework Program (FP7/2007-2013 Grant Agreement no. 279954) ERC-StG ``EDECS''.
\end{acknowledgments}

\end{document}